\documentclass[12pt]{emulateapj}
\usepackage{amssymb}
\usepackage{epsfig}

\newcommand{\etal}{et~al.~}
\newcommand\tna{\,\tablenotemark{a}}

\begin{document}


\title{Star Formation Efficiency in the Cool Cores of Galaxy Clusters}

\author{Michael McDonald\altaffilmark{1,3}, Sylvain Veilleux\altaffilmark{1,4}, David S. N. Rupke\altaffilmark{2}, Richard Mushotzky\altaffilmark{1}, and Christopher Reynolds\altaffilmark{1}}

\altaffiltext{1}{Department of Astronomy, University of Maryland, College
  Park, MD 20742}
\altaffiltext{2}{Department of Physics, Rhodes College, Memphis, TN 38112}   
\altaffiltext{3}{Email: mcdonald@astro.umd.edu}
\altaffiltext{4}{Email: veilleux@astro.umd.edu}


\begin{abstract}
We have assembled a sample of high spatial resolution far-UV (Hubble
Space Telescope Advanced Camera for Surveys Solar Blind Channel) and
H$\alpha$ (Maryland-Magellan Tunable Filter) imaging for 15 cool core
galaxy clusters. These data provide a detailed view of the thin,
extended filaments in the cores of these clusters. Based on the ratio
of the far-UV to H$\alpha$ luminosity, the UV spectral energy distribution, and the far-UV and
H$\alpha$ morphology, we conclude that the warm, ionized gas in the
cluster cores is photoionized by massive, young stars in all but a few
(Abell~1991, Abell~2052, Abell~2580) systems. We show that the
extended filaments, when considered separately, appear to be
star-forming in the majority of cases, while the nuclei tend to have
slightly lower far-UV luminosity for a given H$\alpha$ luminosity,
suggesting a harder ionization source or higher extinction. We observe a slight offset in the UV/H$\alpha$ ratio from the expected value for continuous star formation which can be modeled by assuming intrinsic extinction by modest amounts of dust (E(B-V)~$\sim$~0.2), or a 
top-heavy IMF in the extended filaments. The measured star formation
rates vary from $\sim$~0.05~M$_{\odot}$~yr$^{-1}$ in the nuclei of
non-cooling systems, consistent with passive, red ellipticals, to
$\sim$~5~M$_{\odot}$~yr$^{-1}$ in systems with complex, extended,
optical filaments. Comparing the estimates of the star formation rate
based on UV, H$\alpha$ and infrared luminosities to the
spectroscopically-determined X-ray cooling rate suggests a star
formation efficiency of 14$^{+18}_{-8}$\%. This value represents the
time-averaged fraction, by mass, of gas cooling out of the
intracluster medium which turns into stars, and agrees well with
the global fraction of baryons in stars required by simulations to reproduce the stellar mass function for galaxies. This result provides a new constraint
on the efficiency of star formation in accreting systems.
\end{abstract}



\section{Introduction}
The high densities and low temperatures of the intracluster medium
(hereafter ICM) in the cores of some galaxy clusters suggests that
massive amounts (100--1000 M$_{\odot}$~yr$^{-1}$) of cool gas should
be deposited onto the central galaxy. The fact that this gas reservoir
is not observed has been used as prime evidence for feedback-regulated
cooling (see review by Fabian 1994). By invoking feedback, either by
active galactic nuclei (hereafter AGN) (e.g., Guo \etal 2008; Rafferty
\etal 2008; Conroy \etal 2008), mergers (e.g., G\'{o}mez \etal 2002;
ZuHone 2010), conduction (e.g., Fabian \etal 2002; Voigt \etal 2004),
or some other mechanism, theoretical models can greatly reduce the
efficiency of ICM cooling, producing a better match with what is
observed in high resolution X-ray grating spectra of cool cores
(0--100 M$_{\odot}$ yr$^{-1}$, Peterson \etal 2003).  However,
these modest cooling flows had remained unaccounted for at low
temperatures until only recently.

The presence of warm, ionized gas in the form of H$\alpha$ emitting
filaments has been observed in the cores of several cooling flow
clusters to date (e.g., Hu \etal 1985, Heckman \etal 1989, Crawford
\etal 1999, Jaffe \etal 2005, Hatch \etal 2007). More recently, it has
been shown by McDonald \etal (2010, 2011; herafter M+10 and M+11,
respectively) that this emission is intimately linked to the cooling
ICM and may be the result of cooling instabilities. However, while it
is possible that the warm gas may be a byproduct of ICM cooling, the
source of ionization in this gas remains a mystery. A wide variety of
ionization mechanisms are viable in the cores of clusters (see
Crawford \etal 2005 for a review), the least exotic of which may be
photoionization by massive, young stars.

\begin{table*}[htb]
\caption{Sample of 15 cooling flow clusters with MMTF H$\alpha$ and
HST FUV imaging}
\begin{center}
\begin{tabular}{c c c c c c c c}
\hline\hline
Name & RA & Dec & z & E(B-V) & \.{M} & F$_{1.4}$ & Proposal No. \\
(1) & (2) & (3) & (4) & (5) & (6) & (7) & (8) \\
\hline
Abell 0970\tna & 10h17m25.7s & -10d41m20.3s & 0.0587 & 0.055 & -- & $<$ 2.5 & 11980\\
Abell 1644     & 12h57m11.6s & -17d24m33.9s & 0.0475 & 0.069 & 3.2 & 98.4 & 11980\\
Abell 1650     & 12h58m41.5s & -01d45m41.1s & 0.0846 & 0.017 & 0.0 & $<$ 2.5 & 11980\\
Abell 1795     & 13h48m52.5s & +26d35m33.9s & 0.0625 & 0.013 & 7.8 & 924.5 & 11980, 11681\\
Abell 1837     & 14h01m36.4s & -11d07m43.2s & 0.0691 & 0.058 & 0.0 & 4.8 & 11980\\
Abell 1991     & 14h54m31.5s & +18d38m32.4s & 0.0587 & 0.025 & 14.6 & 39.0 & 11980\\
Abell 2029     & 15h10m56.1s & +05d44m41.8s & 0.0773 & 0.040 & 3.4 & 527.8 & 11980\\
Abell 2052     & 15h16m44.5s & +07d01m18.2s & 0.0345 & 0.037 & 2.6 & 5499.3 & 11980\\
Abell 2142     & 15h58m20.0s & +27d14m00.4s & 0.0904 & 0.044 & 1.2 & $<$ 2.5 & 11980\\
Abell 2151     & 16h04m35.8s & +17d43m17.8s & 0.0352 & 0.043 & 8.4 & 2.4 & 11980\\
Abell 2580\tna     & 23h21m26.3s & -23d12m27.8s & 0.0890 & 0.024 & -- & 46.4 & 11980\\
Abell 2597     & 23h25m19.7s  & -12d07m27.1s  & 0.0830 & 0.030 &  9.5 & 1874.6 & 11131\\
Abell 4059     & 23h57m00.7s & -34d45m32.7s & 0.0475 & 0.015 & 0.7 & 1284.7 & 11980\\
Ophiuchus      & 17h12m27.7s & -23d22m10.4s & 0.0285 & 0.588 & 0.0 & 28.8 & 11980\\
WBL 360-03\tna & 11h49m35.4s & -03d29m17.0s & 0.0274 & 0.028 & -- & $<$ 2.5 & 11980\\
\hline
\\
\end{tabular}
\end{center}
\em{}
\vskip -0.2 in
(1): Cluster name, (2--4): NED RA, Dec, redshift of BCG (\url{http://nedwww.ipac.caltech.edu}), (5): Reddening due to Galactic extinction from Schlegel \etal (1998), (6): Spectroscopically-determined X-ray cooling rates (M$_{\odot}$ yr$^{-1}$) from McDonald \etal (2010), (7): 1.4 GHz radio flux (mJy) from NVSS (\url{http://www.cv.nrao.edu/nvss/}) (8) HST proposal number for FUV data. Proposal PIs are W. Jaffe (\#11131), W. Sparks (\#11681), S. Veilleux (\#11980). \\$^a$: No available \emph{Chandra} data.
\label{sample}
\end{table*}

The identification of star-forming regions in cool core clusters has a rich history in the literature. Early on, it was noted by several groups that brightest cluster galaxies (hereafter BCGs) in cool core clusters have higher star formation rates than non-cool core BCGs (Johnstone \etal 1987; Romanishin \etal 1987; McNamara and O'Connell 1989; Allen 1995; Cardiel \etal 1995). These studies all found evidence for significant amounts of star formation in cool cores, but the measured star formation rates were orders of magnitudes smaller than the X-ray cooling rates (e.g. McNamara and O'Connell 1989).  In recent history, two separate advances have brought these measurements closer together. First, as mentioned earlier in this section, the X-ray spectroscopically-determined cooling rates are roughly an order of magnitude lower than the classically-determined values based on the soft X-ray luminosity. Secondly, large surveys in the UV (e.g., Rafferty \etal 2006; Hicks \etal 2010), optical (e.g., Crawford \etal 1999; Edwards \etal 2007; Bildfell \etal 2008; McDonald \etal 2010), mid-IR (e.g., Hansen \etal 2000; Egami \etal 2006; Quillen \etal 2008; O'Dea \etal 2008; hereafter MIR), and sub-mm (e.g., Edge 2001; Salom\'e and Combes 2003) have allowed a much more detailed picture of star formation in BCGs. The typical star formation rates of $\sim$ 1--10 M$_{\odot}$ yr$^{-1}$ (O'Dea \etal 2008) imply that gas at temperatures of $\sim10^{6-7}$~K is being
continuously converted into stars with an efficiency on the order of $\sim$
10\%. The fact that most of these studies consider the
\emph{integrated} SF rates makes it difficult to determine the exact
role of young stars in ionizing the extended warm gas observed at
H$\alpha$, since the two may not be spatially coincident or the
measurements may be contaminated by the inclusion of a central AGN.

In order to understand both the role of star formation in ionizing the
warm gas and the efficiency with which the cooling ICM is converted
into stars, we have conducted a high spatial resolution far-UV survey
of BCGs in cooling and non-cooling clusters. We describe the
collection and analysis of the data from this survey in \S2. In \S3 we
desribe the results of this survey, while in \S4 we discuss the
implications of these results in the context of our previous work
(M+10,M+11). Finally, in \S5 we summarize our findings and discuss any outstanding questions. 
Throughout this paper, we assume the following cosmological
values: H$_0$ = 73 km s$^{-1}$ Mpc$^{-1}$, $\Omega_{matter}$ = 0.27,
$\Omega_{vacuum}$ = 0.73.


\section{Data Collection and Analysis}

To study the far-UV (hereafter FUV) emission in cluster cores, we
selected 15 galaxy clusters from the larger samples of M+10 and
M+11, which have deep, high spatial resolution (FWHM $\sim$ 0.6$^{\prime\prime}$) H$\alpha$ imaging from
the Maryland-Magellan Tunable Filter (hereafter MMTF; Veilleux \etal
2010) on the Baade 6.5-m telescope at Las Campanas
Observatory. Additionally, most of these systems have deep Chandra
X-ray Observatory (hereafter CXO) spectroscopic imaging, as well as
Two-Micron All Sky Survey (hereafter 2MASS; Skrutskie \etal 2006) and
NRAO VLA Sky Survey (hereafter NVSS; Condon \etal 1998) fluxes. This
broad energy coverage provides an excellent complement to a FUV
survey, allowing for the source of emission to be carefully
identified. A summary of these 15 clusters can be found in Table 1.
For further information about the reduction and analysis of the
H$\alpha$ and X-ray data, see M+10.

FUV imaging was acquired using the Advanced Camera for Surveys Solar
Blind Channel (hereafter ACS/SBC) on the Hubble Space Telescope
(hereafter HST) in both the F140LP and F150LP bandpasses whenever
possible, with a total exposure time of $\sim$ 1200s each (PID
\#11980, PI Veilleux). The pointings were chosen, based on the results
of our MMTF survey, to include all of the H$\alpha$ emission in the
field of view.  Exposures with multiple filters are required to
properly remove the known ACS/SBC red leak, which has a non-negligible
contribution due to the fact that the underlying BCG is very luminous
and red. Since the aforementioned filters are long-pass filters, they
have nearly identical throughputs at longer wavelengths. Thus, by
subtracting the F150LP exposure from the F140LP exposure we can
effectively remove the red leak and consider only a relatively small
range in wavelength, from 1400\AA--1500\AA. Due to the small bandpass, it is possible for line emission to dominate the observed flux -- we investigate this possibility in \S4. We have carried out this
subtraction for 13/15 of the BCGs in our sample which have both F140LP
and F150LP imaging. For Abell~1795 and Abell~2597, we are unable to
remove the red leak contribution due to the lack of paired exposures,
but we point out that, conveniently, these two systems have the
brightest FUV flux in our sample and, thus, are largely unaffected by
the inclusion of a small amount of non-FUV flux.  Following the
red-leak subtraction, we also bin the images $8\times8$ and smooth the images with a 1.5 pixel smoothing radius, yielding
matching spatial resolution at FUV and H$\alpha$. This process is also
necessary in order to increase the signal-to-noise of the FUV image,
allowing us to identify interesting morphological features. The final pixel scale for both the H$\alpha$ and FUV images is 0.2$^{\prime\prime}$/pixel.

All FUV and H$\alpha$ fluxes were corrected for Galactic extinction
following Cardelli \etal (1989) using reddening estimates from
Schlegel \etal (1998).

\section{Results}

\begin{figure*}
\begin{center}
\includegraphics[width=0.9\textwidth]{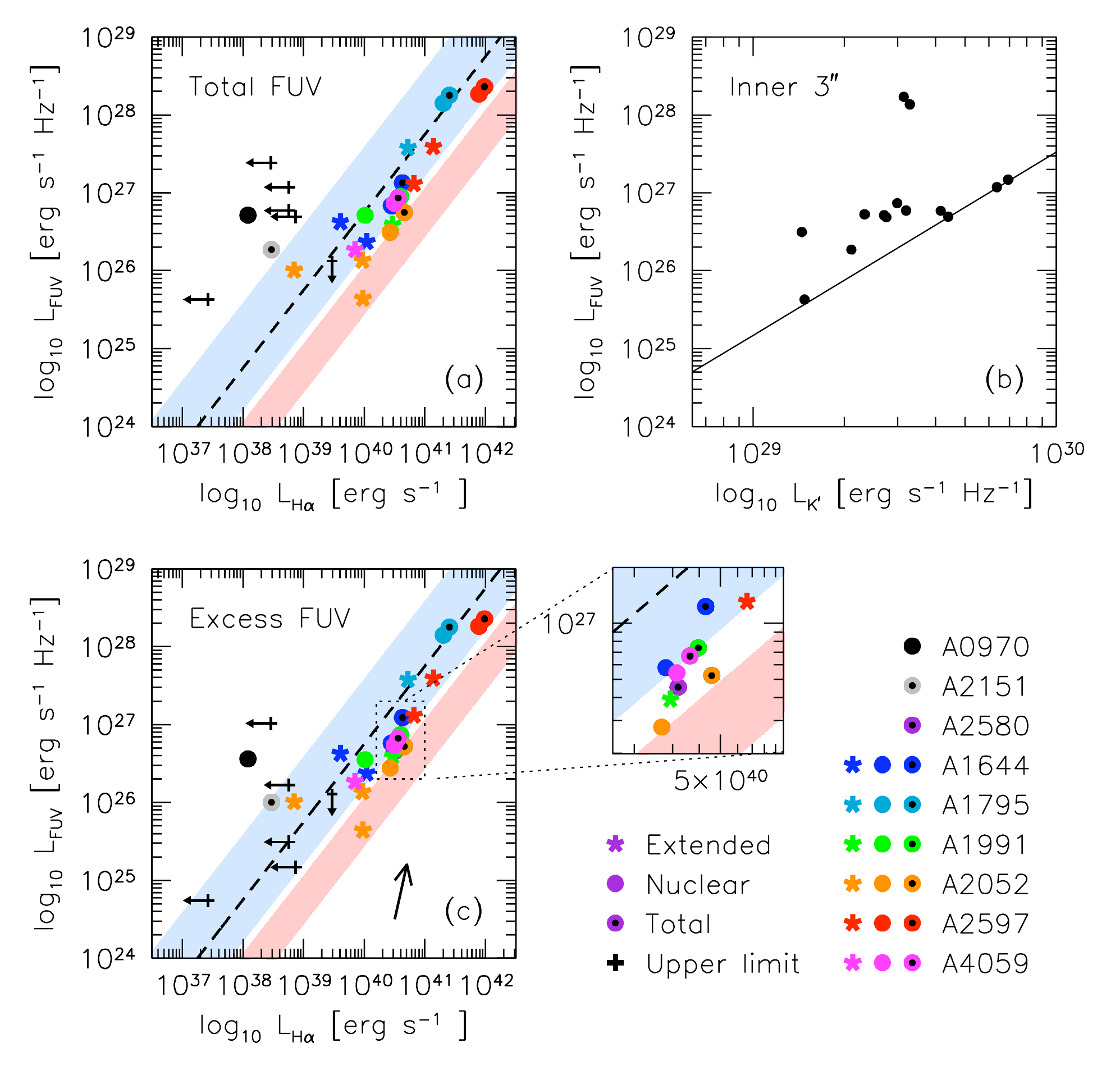}
\caption{(a) FUV versus H$\alpha$ luminosity for 15 BCGs in our
sample. For systems with extended FUV or H$\alpha$ emission, we
separate the emission into extended, nuclear (inner
3$^{\prime\prime}$) and entire systems (see Appendix). The dashed line
represents the relation between H$\alpha$ and FUV luminosity for SF
regions, as defined by Kennicutt (1998), the shaded blue region
represents the expected FUV/H$\alpha$ ratio for continuous SF covering
the full range of IMFs and metallicities from Starburst99 (Leitherer \etal
1999), while the shaded red region defines the expected FUV/H$\alpha$
ratio for fast shocks (Dopita \& Sutherland 1996). The legend in the
lower right describes the different point type/colors. (b) FUV versus
K$^{\prime}$-band luminosity in the central 3$^{\prime\prime}$ of the
BCG. The solid line represents our estimate of the contribution to the
FUV luminosity from old, horizontal-branch stars. (c) Similar to panel
(a), but with the contribution from old stars removed from the nuclear
and total regions. The correlation between FUV and H$\alpha$
luminosity is much more significant, suggesting that the majority of
the observed H$\alpha$ emission may be due to photoionization by young
stars. The extended and nuclear emission in Abell~2052 and Abell~2580, as well as the extended emission in Abell~1991 appears to be consistent with heating by shocks, while
the remaining systems are likely star-forming. The arrow in the lower
right corner represents the magnitude of the intrinsic extinction
correction for E(B-V)=0.2.}
\label{uvha}
\end{center}
\end{figure*}

In the Appendix, we show the stellar continuum, H$\alpha$ and FUV
images for each of the 15 BCGs in our sample. At a glance there does
not appear to be consistent agreement between the H$\alpha$ and FUV
morphologies.  We observe systems having H$\alpha$ filaments without
accompanying FUV emission (Abell~1991, Abell~2052), systems with
complex FUV emission without accompanying H$\alpha$ (Abell~1837,
Abell~2029), and systems with coincident H$\alpha$ and FUV extended
emission (Abell~1644, Abell~1795). Thus, it is obvious that a single
explanation (e.g., star formation) is unable to account for the
variety of FUV and H$\alpha$ emission that we observe.

As we did with the H$\alpha$ emission in M+10, the FUV morphology can
be classified as either nuclear or extended. We find, in the FUV, 7/15
systems have extended emission, 5/15 have nuclear emission, while 3/15
have no emission at all.  In order to quantitatively examine both the
nuclear and extended emission, we extract FUV and H$\alpha$ fluxes in
several regions, as shown in the Appendix.

\begin{figure}
\begin{center}
\includegraphics[width=0.48\textwidth]{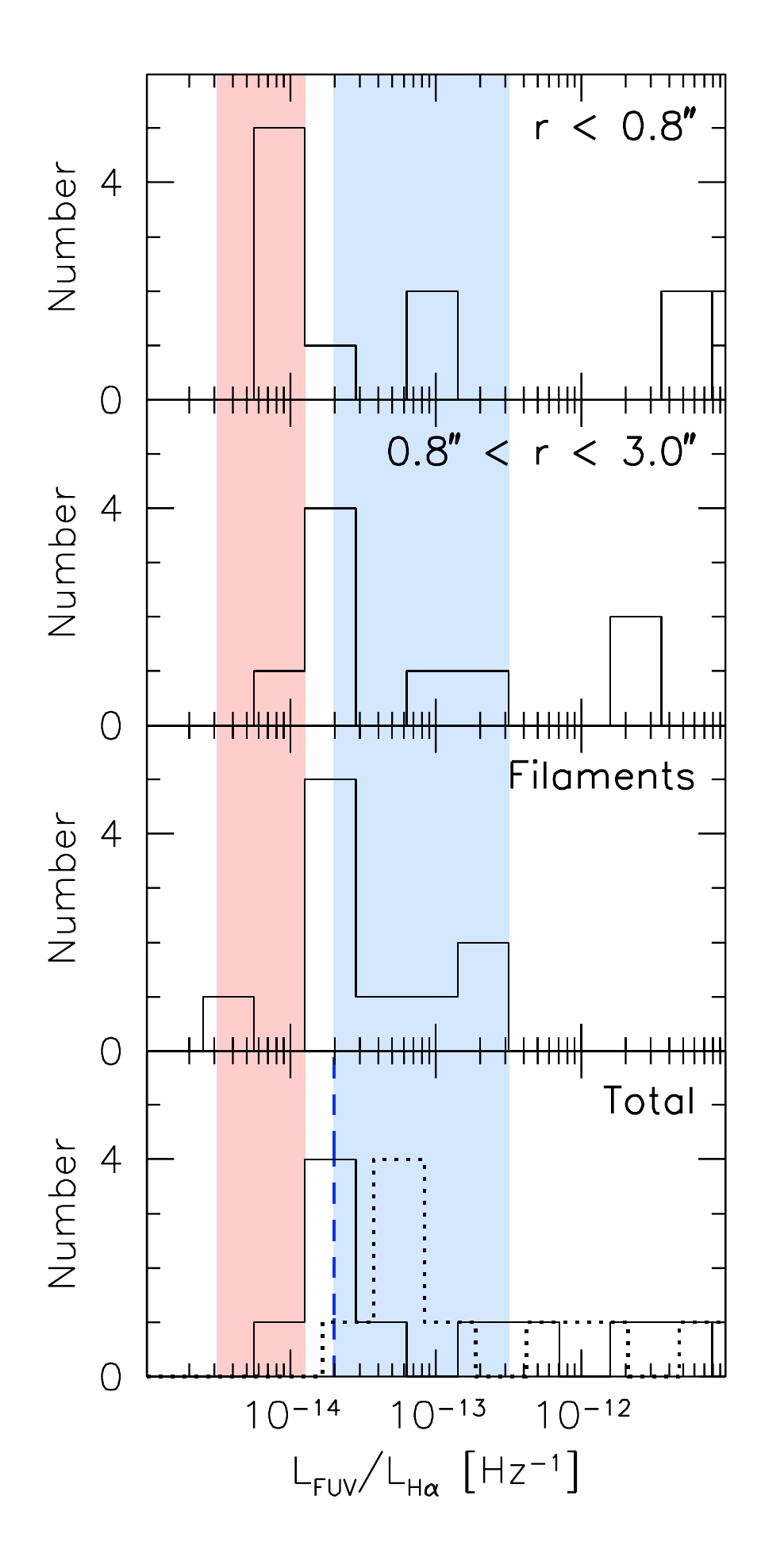}
\caption{Distribution of the FUV to H$\alpha$ luminosity ratios
for the 10 clusters in our sample with high S/N detections in either
the FUV or H$\alpha$. Each panel considers a different region, from
top to bottom: the nucleus, the central 3$^{\prime\prime}$, excluding
the nucleus, the filaments, and the entire system. The red vertical
band represents the region consistent with shocks (Dopita \&
Sutherland 1996), while the blue band represents the region consistent
with continuous star formation (Leitherer \etal 1999). In the nucleus,
the majority of the observed H$\alpha$ emission is consistent with
having been shock-heated, while in the outer regions, including
filaments, the FUV/H$\alpha$ ratio lies between the two highlighted
regions. In the bottom panel, the dotted histogram shows how the
distribution would be altered if we corrected the fluxes for an i
intrinsic extinction of E(B-V)=0.2 (dust screen model), while the blue
dashed line represents the expected value for O8V stars (symbolic of a
top-heavy IMF).}
\label{uvha_hists}
\end{center}
\end{figure}

In Figure \ref{uvha}a, we show the correlation between the FUV and
H$\alpha$ luminosity for the regions identified in the Appendix.  We
find a significant amount of FUV emission in all 5 of the systems for
which we do not detect any H$\alpha$ emission. Additionally, we see
that at least 3 systems are consistent with being shock-heated (Dopita
\& Sutherland 1996) -- a point we will return to later in this
section.

As discussed by Hicks \etal (2010), a significant fraction of the FUV
emission may be due to old stellar populations (e.g., horizontal
branch stars) in the BCG. To proceed, we must isolate the FUV excess
due to young, star-forming regions. In order to remove the
contribution from old stars, we consider the inner 3$^{\prime\prime}$
and plot the K-band (from 2MASS; Skrutskie \etal 2006) versus the FUV
luminosity (Figure \ref{uvha}b). Hicks \etal (2010) show that the FUV
luminosity from old stars is highly concentrated in the central
region, thus removing this contribution in the inner region will act
as a suitable first-order correction. In order to calibrate this correction for our sample, which lacks a control sample of confirmed non-star-forming galaxies, we opt to fit a line which is chosen to pass through the four
points with the lowest $L_{FUV}$/$L_{K^{\prime}}$ ratio. We make the assumption that these four galaxies with the lowest $L_{FUV}/L_{K^{\prime}}$ ratio are non-star-forming, which is supported by non-detections at H$\alpha$.  The equation for this relation is: log$_{10}(L_{FUV,3^{\prime\prime}}) = 2.35$log$_{10}(L_{K^{\prime},3^{\prime\prime}}) - 42.98$. The fact that four points with non-detections at H$\alpha$ lie neatly along the same line 
suggests that this correction is meaningful.

Figure \ref{uvha}c shows the FUV excess due to young stars versus the
H$\alpha$ luminosity for the total, nuclear and extended regions in
our complete sample of BCGs. With the contribution from old stellar
populations removed, we find a tight correlation between $L_{FUV}$ and
$L_{H\alpha}$ over four orders of magnitude. The majority of systems
in our sample are consistent with the continuous star formation
scenario (Kennicutt 1998, Leitherer \etal 1999), suggesting that much
of the warm gas found in cluster cores may be photoionized by young
stars. Two systems, Abell~0970 and Abell~2029 have anomolously high
FUV/H$\alpha$ ratios, suggesting that star formation may be proceeding
in bursts. As a starburst ages, the UV/H$\alpha$ ratio will climb quickly due to the massive stars dying first. This means that, by 10 Myr after the burst, the UV/H$\alpha$ ratio can already be an order of magnitude higher than the expected value for continuous star formation (see M+10 for further discussion). We find that the filaments in
Abell~1991 and Abell~2052 are consistent with being
heated by fast shocks, along with the nuclei of Abell~2052 and Abell~2580. In the case of Abell~2052, there exist high quality radio and X-ray maps which show that the observed H$\alpha$ emission is coincident with the inner edge of a radio-blow bubble. In Abell~1991, the H$\alpha$ morphology is reminiscent of a bow shock, and is spatially coincident with a soft X-ray blob which is offset from the cluster core. Much of
the FUV and H$\alpha$ data is clustered between the regions depicting continuous star
formation and shock heating, as shown in the zoomed-in portion of
Figure \ref{uvha}. These regions may indeed be heated by a combination
of processes, or they may simply be reddened due to intrinsic
extinction. Based on their FUV/H$\alpha$ ratios, H$\alpha$
morphology, disrupted X-ray morphology, and high radio luminosity, we propose that the optical emission in Abell~1991, Abell~2052, and Abell~2580 is the product of shock heating, while the
remaining 12 systems are experiencing continuous or burst-like star
formation. We will return to this classification in the
discussion. The SF rates that we measure in the systems with nuclear
emission only range from 0.01--0.1 M$_{\odot}$~yr$^{-1}$, which are
typical of normal red-sequence ellipticals (Kennicutt 1998). For the
systems with extended emission, excluding those that are obviously
shock-heated, the measured star formation rates range from
0.1--5~M$_{\odot}$~yr$^{-1}$ which is similar to the rates of
0.008--3.6 M$_{\odot}$ ~yr$^{-1}$ observed in ``blue early-type
galaxies'' (Wei \etal 2010).

\begin{figure*}
\begin{center}
\begin{tabular}{c c}
\includegraphics[width=0.49\textwidth]{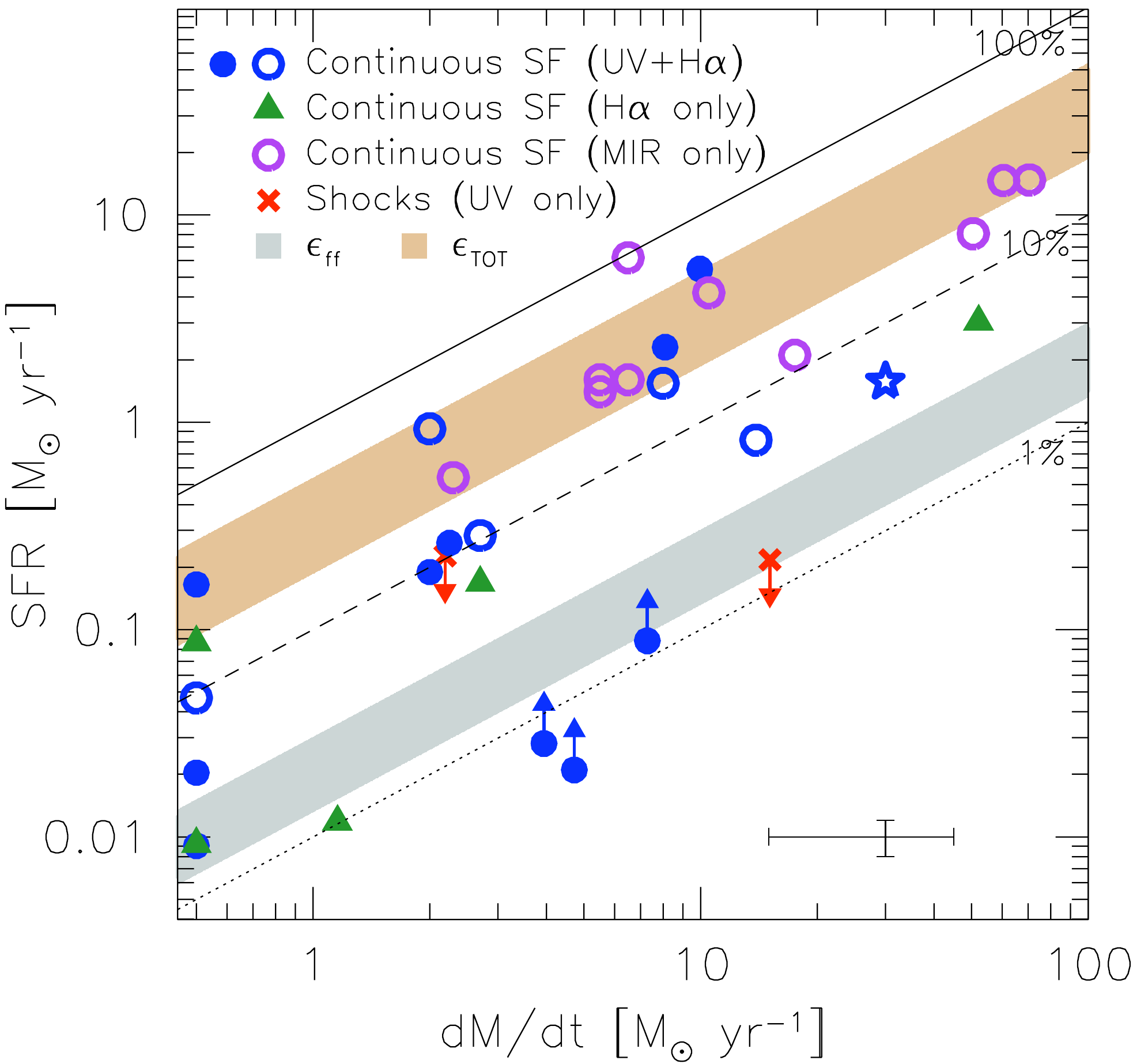} &
\includegraphics[width=0.47\textwidth]{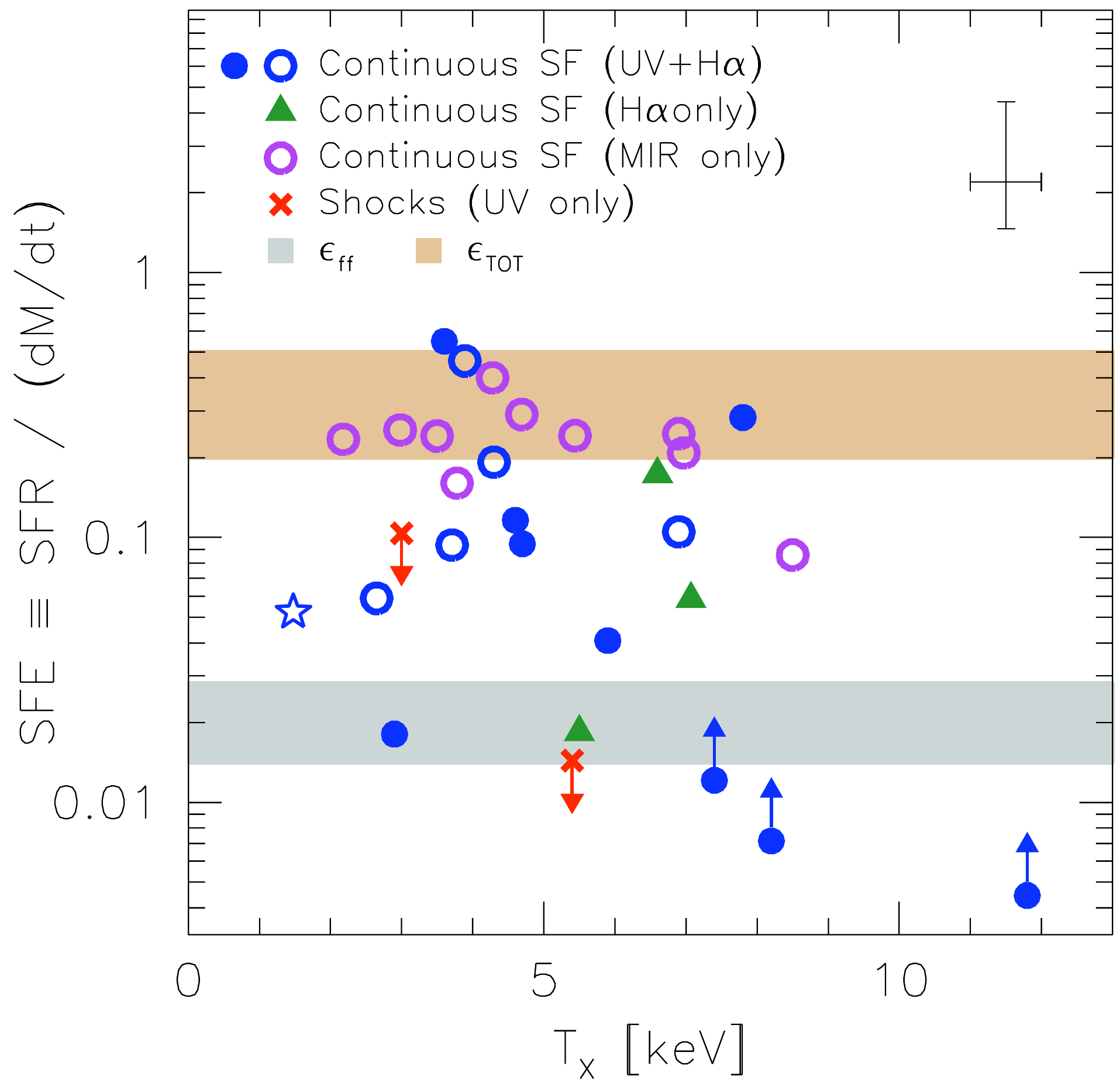}\\
\end{tabular}
\caption{Left: SF rate (SFR) versus X-ray cooling rate (dM/dt) for 32
galaxy clusters. The filled blue circles (10) and red crosses (2) correspond to 
clusters identified as star-forming and shock-heated from Figure
\ref{uvha}, respectively. One of the shock-heated systems, Abell~2580, does not have available data from the CXO archives. The open blue circles (5) refer to clusters with MMTF H$\alpha$ imaging and archival GALEX near-UV (NUV) data from M+10. The green triangles (5) are clusters with deep
H$\alpha$ imaging from M+10, M+11 -- we assume this emission is due to
star formation. Star formation rates are computed from NUV, FUV, and
H$\alpha$ luminosities based on the prescriptions in Kennicutt
(1998). We also show data from O'Dea \etal (2008) as open purple circles (10), with SF
rates based on mid-IR (MIR, 24$\mu m$) {\em Spitzer} data and dM/dt computed in a similar
fashion to us. The blue star represents the Perseus cluster (Conselice
\etal 2001, Sanders \etal 2004). Diagonal lines refer to efficiencies
of 1\%, 10\% and 100\%, while the shaded regions refer to the typical
SF efficiency in molecular clouds per free fall time (gray; Krumholz
\& McKee 2005) and over the lifetime of the cloud (tan; Kroupa \etal 2001; Lada \& Lada 2003). Typical
errors (20\% in SFR, 50\% in dM/dt) are shown in the bottom right
corner. Right: SF efficiency (SFR per dM/dt) as a function of
cluster X-ray temperature from the ACCEPT sample of Cavagnolo \etal
(2009). Typical errors are shown in the upper right corner. We note
that the error in SF efficiency is dominated by our uncertainty in the
X-ray cooling rate, dM/dt. The highest-temperature point is the
Ophiuchus cluster, which lies close to the Galactic plane and, thus,
suffers from heavy extinction. While we do correct for Galactic extinction, the large column density will amplify any uncertainty in the measurement of $N_H$ or in the extinction curves applied.}
\label{sfr}
\end{center}
\end{figure*}

In Figure \ref{uvha_hists} we present the distribution of the
FUV/H$\alpha$ ratio in various regions for the 10 systems with detections ($>$1-$\sigma$) at either H$\alpha$ or FUV. In the innermost region
($r<0.8^{\prime\prime}$), the warm gas appears to be shock-heated in
60\% of systems -- these shocked nuclei may be associated with AGN-driven outflows. Due to the small radial extent of
this bin, the 2MASS data is of insufficient spatial resolution to
remove any contribution from old stars. Thus, these FUV/H$\alpha$
ratios are upper limits. Of the remaining 4 systems, 2 are consistent
with continuous star formation or a young starburst, while the remaining
two are consistent with an aged starburst. At larger radius
($0.8^{\prime\prime}<r<3.0^{\prime\prime}$) the FUV/H$\alpha$ ratio is
slightly larger, with the distribution peaking in between the regions
describing shocks and star formation (see inset of Figure
\ref{sfr}). These data have had the contribution from old stellar populations removed, as described earlier in this section. The FUV/H$\alpha$ ratio at this radius is similar to what
we observe in the filaments, as is seen in the third panel of Figure
\ref{uvha_hists}. The fact that the distribution of FUV/H$\alpha$
peaks between the values for shocks and star formation supports a
number of scenarios, including a mixture of the two processes, dusty
star formation and star formation with an IMF skewed towards high-mass
stars (see right-most panel of Figure \ref{uvha_hists}). We will
investigate these scenarios in \S4.

In M+10 and M+11, we showed that the H$\alpha$ emission observed in
the cool cores of galaxy clusters is intimately linked to the cooling
ICM. In general, the thin, extended filaments observed in many of
these clusters are found in regions where the ICM is cooling most
rapidly, suggesting that this warm gas may be a byproduct of the
ongoing cooling. If this is the case, it is relevant to ask what
fraction of the cooling ICM is turning into stars. We address this
question in Figure \ref{sfr} by comparing the star formation rate with
the X-ray cooling rate (dM/dt) for 32 galaxy clusters. In order to
compute the star formation rate, we use the prescriptions in Kennicutt
(1998). For the systems observed with HST, we use the average of the
FUV- and H$\alpha$-determined star formation rates (filled blue
circles). For an additional 10 clusters from M+10 and M+11 we make use of
archival GALEX data for 5 clusters (open blue circles) and assume that
both the UV and H$\alpha$ emisison trace star forming regions. These data have also had the contribution from old stellar populations removed, as described in M+10. In the
remaining 5 cases where there is no accompanying UV data, we assume
that the H$\alpha$ emission is the result of photoionization by young
stars, and convert the H$\alpha$ luminosity into a continuous star
formation rate (green triangles). 
For the three shock-heated systems (red crosses) mentioned above, we determine the SF rate based on the FUV luminosity alone -- this represents an upper limit on the amount of continuous star formation. Finally, we
also include 10 clusters from the sample of O'Dea \etal (2008), who
compute SF rates based on {\em Spitzer} data, and dM/dt in a similar
manner to us (open purple circles). We note that the SFRs derived
from FUV data are systematically lower than those derived from 24$\mu
m$ {\em Spitzer} data. This is most likely due to our lack of an
intrinsic extinction correction for these FUV data.

We find that the ``efficiency'' of star formation, defined as the
current ratio of stars formed to gas cooling out of the ICM, can range
from 1\% to 50\% over the full sample. This spread is independent of
whether we only consider systems classified as star-forming based on
their FUV/H$\alpha$ ratios in Figure \ref{uvha}.  For the majority of
the 32 clusters in Figure \ref{sfr}, there is a tight correlation
between SF rate and dM/dt with a typical efficiency between
10--50\%. In the right panel of Figure \ref{sfr}, we show that
the efficiency is nearly independent of the cluster temperature, with
only a weak dependence which is primarily driven by highly extincted
systems.  This suggests that the temperature of the surrounding ICM does not hamper the BCG's ability to form stars. While not shown here, we also investigated the distribution of star formation efficiencies with the central entropy, $K_0$, from Cavagnolo \etal (2009) and, similarly to $T_X$, find no correlation. In the following section we will discuss possible
interpretations of this efficiency measure.


\section{Discussion}
\subsection{Star Formation as an Ionization Source}
\begin{figure*}[htb]
\begin{center}
\begin{tabular}{c c}
\includegraphics[width=0.45\textwidth]{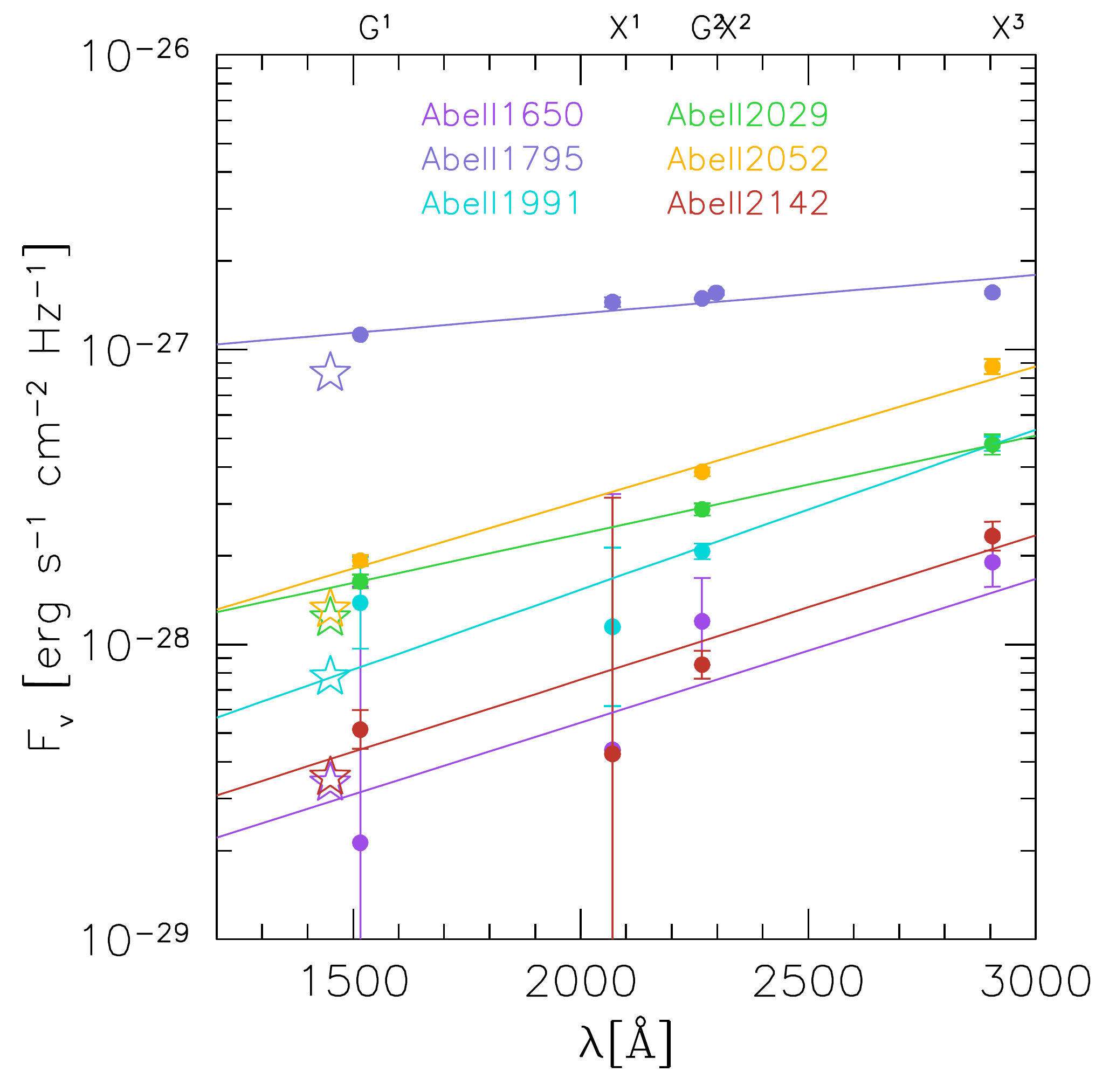} &
\includegraphics[width=0.445\textwidth]{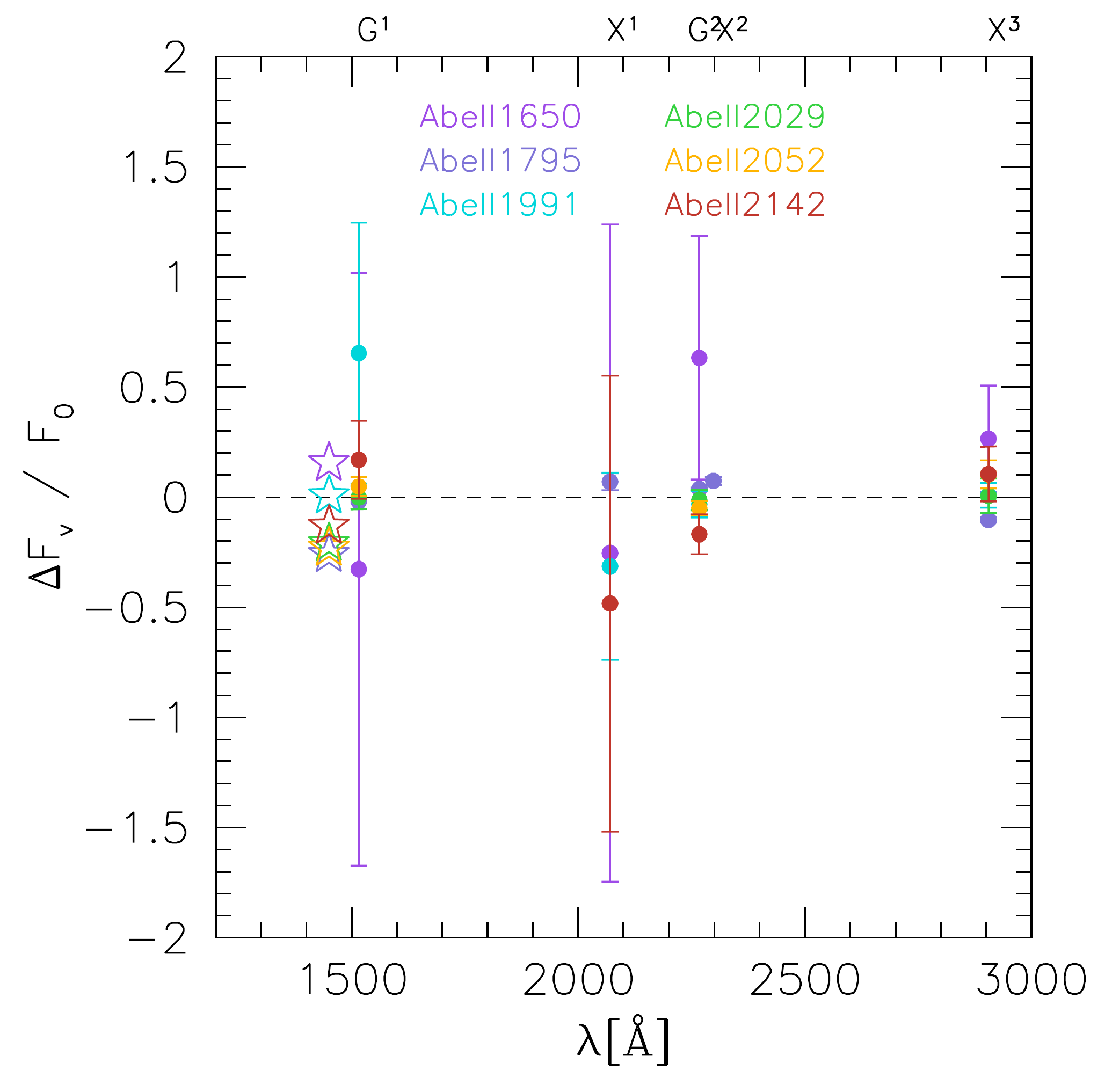} \\
\end{tabular}
\caption{ Left panel: UV spectral energy distributions (SEDs) for 6 cluster cores in our sample using data from GALEX and XMM-OM (M+10). All fluxes have been corrected for Galactic extinction. The two \emph{GALEX} filters are denoted by G$^1$ and
G$^2$, while the three \emph{XMM-OM} filters are denoted by X$^1$,
X$^2$ and X$^3$. The colored lines represent a fit to the GALEX+XMM-OM data, while the stars represent the FUV flux that we measure with HST ACS/SBC. Right panel: Residuals from the continuum fits shown in the left panel. Our new HST FUV data are consistent with the UV continuum fit, showing little evidence for additional line emission.}
\label{sed}
\end{center}
\end{figure*}

In Crawford \etal (2005), a variety of ionization sources are
discussed which could produce the observed H$\alpha$ emission in the
cool cores of galaxy clusters. The purpose of this HST survey was to
investigate one of the most plausible ionization sources:
photoionization by young stars. In Figures \ref{uvha} and
\ref{uvha_hists} we showed that, once the contribution to the FUV
emission from old stellar populations is removed, the majority of the
H$\alpha$ and FUV emission that we observe in cluster cores is roughly
consistent with the star formation scenario. Based on the
FUV/H$\alpha$ ratios, we identify three different types of system:

\begin{enumerate}

\item FUV/H$\alpha$ $\gtrsim$ $10^{-12}$ Hz$^{-1}$. Suggests a
starburst that has aged by at least 10 Myr. Two systems, Abell~0970
and Abell~2029, fulfill this criteria, while several others may fall
into this category if their H$\alpha$ luminosity is significantly less
than the measured upper limits.

\item FUV/H$\alpha$ $\sim$ $10^{-13}$ Hz$^{-1}$. The FUV/H$\alpha$
ratios of these systems are consistent with continuous star formation or
a recent (0--5 Gyr ago) burst of star formation. The filaments in
Abell~1644, Abell~1795, Abell~2597, and part of Abell~2052, along with
the nuclei of Abell~1795, Abell~1991, Abell~2151, and Abell~2597
appear to be star-forming.

\item FUV/H$\alpha$ $\lesssim$ $10^{-14}$ Hz$^{-1}$. Suggests heating
by fast shocks or some other source of hard ionization (e.g., cosmic
rays, AGN). The filaments of Abell~2052 and Abell~1991, and the nuclei of
Abell~2052 and Abell~2580 have FUV/H$\alpha$ ratios which are
consistent with this picture, in the absence of internal reddening.

\end{enumerate}

Figures \ref{uvha} (inset) and \ref{uvha_hists} show that a large
fraction of the systems which we observe fall between the regions
describing shock heating and star formation. However, these data have
not been corrected for intrinsic reddening due to dust. Correcting for
a very modest reddening (E(B-V) $\sim$ 0.2) would boost the FUV
luminosity of these systems such that the FUV/H$\alpha$ ratio is
consistent with star formation (see Figure \ref{uvha}c and the lower
panel of Figure \ref{uvha_hists}). Unfortunately, the amount of
reddening in the filaments and nuclei of these systems is currently
not well constrained for very many systems, but typical values of
E(B-V) can range from 0--0.4 in the cores of galaxy clusters (Crawford
\etal 1999). In the case of Abell~2052, for which we measure FUV/H$\alpha$ ratios indicative of shock-heating \emph{and} have an estimate of the amount of intrinsic reddening from Crawford ($E(B-V)=0.22^{+0.36}_{-0.21}$), we can investigate whether correcting for this extinction would provide FUV/H$\alpha$ ratios consistent with star-forming regions. Assuming a simple dust-screen model, correcting for a reddening of $E(B-V)=0.22$ would transform a FUV/H$\alpha$ ratio of $4.4\times10^{-15} Hz^{-1}$ in the filaments of Abell~2052 to $1.2\times10^{-14} Hz^{-1}$, which is consistent with the upper limit for shock-heated systems (see Figure \ref{uvha_hists}). However if, contrary to expectations, the filaments have a slightly higher reddening than the nucleus,  the FUV/H$\alpha$ ratio may be even higher. Thus, it is certainly possible that the systems which we classify as shock-heated, or those which have ambiguous FUV/H$\alpha$ ratios, may in fact be highly-obscurred star-forming systems. We will address this possibility in significantly more detail in an upcoming paper which will include long-slit spectroscopy of the H$\alpha$ filaments providing, for the first time, reddening estimates away from the nucleus in these systems.

An alternative explanation for the intermediate
FUV/H$\alpha$ ratios is that the IMF in the filaments is top-heavy
($\alpha \ll 2.35$). Again, the lower panel of Figure \ref{uvha_hists}
shows that the peak of the FUV/H$\alpha$ distribution is consistent
with the value expected for O8V stars. There is a substantial amount
of literature providing evidence for a top-heavy IMF in various environments including the Galactic center (Maness \etal 2007) and disturbed galaxies (Habergham \etal 2010). Thus, regardless of whether there is a small amount
of dust or a slightly altered IMF, we suspect that the majority of the
systems with intermediate FUV/H$\alpha$ ratios are in fact
star-forming, with the exception of Abell~1991, Abell~2052, and
Abell~2580, which have low FUV/H$\alpha$ ratios \emph{and}
morphologies which resemble bow-shocks and/or jets.

Due to our use of the F150LP filter to remove red leak
contamination, we are considering only a very small wavelength range
from 1400--1500\AA. In this region, there may be line emission from
[OIV] and various ionization states of sulfur due to gas cooling at
$\sim 10^5$ K. In order to establish that we are indeed observing
continuum emission from young stars, we have computed UV spectral
energy distributions (SEDs) for 6 BCGs which have deep GALEX, XMM-OM,
and HST UV data. These data are presented in the left panel of Figure
\ref{sed}. We see that, in general, the UV SED follows a powerlaw over
the range of 1500--3000\AA. The new HST data, depicted as colored
stars in this plot, agree well with the extrapolation of the continuum
to shorter wavelengths, suggesting that there is very little
contamination from line emission. This also suggests that there is little contribution from a diffuse UV component. This is further emphasized in the
right panel of Figure \ref{sed} where we show the residuals from the
continuum fit for each of the 6 BCGs. Our measured FUV fluxes from
these new HST data are consistent with the measurement of a UV
continuum from archival GALEX and XMM-OM data. 

The idea that massive, young stars may be responsible for heating the
majority of the warm, ionized filaments observed in cool core clusters
is certainly not a new one (see e.g., Hu \etal 1985; Heckman \etal 1989; McNamara and O'Connel 1989). Most recently, O'Dea \etal (2008), Hicks
\etal (2010) and McDonald \etal (2010a) conducted MIR, UV and
H$\alpha$ surveys, respectively, of cool core clusters and found a
strong correlation between the SF rate and the cooling properties of
cluster cores. However, this work extends these findings to include
spatially-resolved SF rates, which the previous studies have been
unable to provide. This allows us to say conclusively that the young
stars and the warm, ionized gas are in close proximity ($\lesssim 1^{\prime\prime}$) in the vast
majority of systems, offering a straightforward explanation for the
heating of these filaments.

\subsection{Star Formation Efficiencies in Cooling Flows}

In \S3, we provide estimates of the efficiency with which the cooling
ICM is converted into stars, assuming that this is indeed the source
of star formation. This assumption is based on the results of M+10 and M+11,
which provided several strong links between the X-ray cooling
properties and the warm, ionized gas. These estimates of star
formation efficiency represent a constraint on the so-called
``accreting box model'' of star formation. The simplified model that
we propose is that the ICM is allowed to cool rapidly in regions where
cooling \emph{locally} dominates over feedback (Sharma \etal
2010). Our estimates of the ICM cooling rate, based on
medium-resolution CXO spectra, are consistent with estimates based on
high-resolution XMM grating spectroscopy by Peterson \etal (2003) for
the five overlapping systems. Once the gas reaches temperatures of
$\sim$10$^{5.5}$K, it can continue to cool rapidly via UV/optical/IR
line emission without producing fluxes that are inconsistent with what
are observed. In the standard way, star formation will proceed once
the gas reaches low enough temperature and high enough
density. Observations by Edge (2001) and Salom\'{e} \& Combes (2003)
show evidence for molecular gas in the cool cores of several galaxy
clusters, consistent with this picture.

\begin{figure}[b]
\begin{center}
\includegraphics[width=0.49\textwidth]{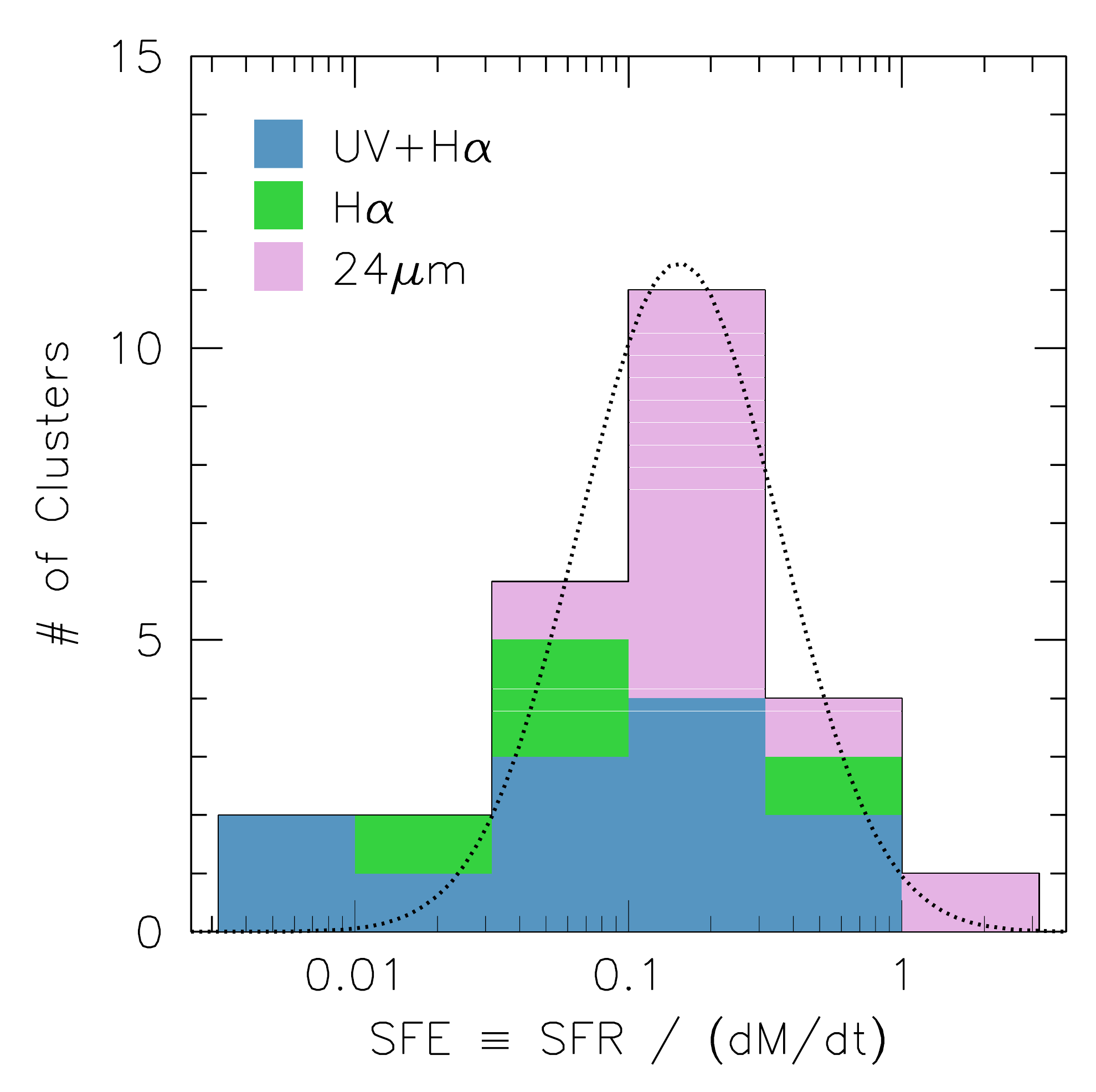}
\caption{Distribution of measured star formation efficiencies from
Figure \ref{sfr} for 26 systems. Systems which are likely shock-heated (Abell~1991,
Abell~2052, Abell~2580) have not been included. The additive
contribution to the total histogram (solid black line) from UV+H$\alpha$, H$\alpha$ and
MIR data are shown in different colors following the color scheme in
Figure \ref{sfr}. The dotted line shows a Gaussian fit to this
histogram, which peaks at $14^{+18}_{-8}$\% efficiency.}
\label{sfre}
\end{center}
\end{figure}

If the above scenario is correct, our estimate of the star
formation efficiency provides a constraint on the fraction of hot gas
that will be converted to stars, assuming a steady inflow of gas. In
Figure \ref{sfre} we provide a histogram of SF efficiencies for all of
the systems in Figure \ref{sfr} with non-zero X-ray cooling rates
(dM/dt). The peak of this distribution is well defined at an
efficiency of $14^{+18}_{-8}$\%, regardless of which SF indicator (UV,
H$\alpha$, MIR) is used.  We note that, while the distribution for UV-
and MIR-determined SF rates both peak at roughly the same value, the
UV-determined SF histogram extends to much lower values. The
low-efficiency tail of this distribution may be an artifact
produced by intrinsic extinction due to dust, to which the UV will be
most sensitive, or may be indicative of a selection bias in the MIR sample. If we measure the peak efficiency based on the subsamples excluding the MIR and MIR+H$\alpha$ (with no accompanying UV) data, we get 10$^{+25}_{-7}$\% and 14$^{+20}_{-8}$\%, respectively. Thus, the peak value of 14\% is not solely driven by the inclusion of MIR data.

The average efficiency of 14$^{+18}_{-8}$\%, based on MIR, H$\alpha$
and UV data, is consistent with the estimates of star formation
efficiency over the lifetime of a typical molecular cloud (20--50\%;
Kroupa \etal 2001, Lada \& Lada 2003).  This large variance in star
formation efficiency may be due to differences in the ICM cooling and
star formation timescales. Naturally, one would expect that there is
some delay between the ICM cooling and the formation of stars, so that
a reservoir of cold gas can accumulate and the formation of stars can
be triggered.  If this is indeed the case, one would expect to observe
cooling-dominated periods (low SFE) followed by periods of strong star
formation (high SFE) once the cold gas reservoir has reached some
critical mass. Over an ensemble of systems, the average SFE is then an
estimate of the time-averaged efficiency of an accreting system in
converting a steady stream of cooling gas into stars. An alternative
explanation for the spread of observed efficiencies is that the source
of feedback is episodic (e.g., AGN). In this scenario, an episode of
strong feedback from the AGN would re-heat the reservoir of cool gas,
severely reducing the potential for star formation. This may indeed be
the case, since two of the three systems with the highest 1.4 GHz
luminosity (Abell~2052, Abell~2597, and Perseus A) have
SFE $\lesssim$ 0.1.

The fact that Figure \ref{sfre} shows a well-defined peak suggests
that the fraction of stars formed in an accreting system is constant 
over long enough timescales. Our estimate
of an average efficiency indicates that, for a steady-state system
accreting hot gas which is then allowed to cool, roughly 4 M$_{\odot}$
of gas will either be re-heated or expelled via winds for every 1
M$_{\odot}$ of stars formed. This fraction of baryons in stars is consistent with the global fraction of $\sim$ 20--30\% required by simulations to reproduce the observed stellar mass function of galaxies (Somerville \etal 2008).
Unlike measurements of SF efficiency for giant molecular clouds,
this estimate does not require the use of a specific timescale, since
we are assuming that stars are forming out of the inflow of hot gas
and that the reservoir for this hot gas is inexhaustible.

\section{Summary and Future Prospects} 
We have assembled a unique set of high spatial resolution far-UV and H$\alpha$ images for 15 cool core galaxy clusters. These data provide
an unprecedented view of the thin, extended filaments in the cores of
galaxy clusters. Based on the ratio of the far-UV to H$\alpha$
luminosity, the UV SED, and the far-UV and H$\alpha$ morphology, we
conclude that the warm, ionized gas in the cluster cores is
photoionized by massive, young stars in all but a few (Abell~1991,
Abell~2052, Abell~2580) systems. We show that the extended filaments,
when considered separately, appear to be forming stars in the majority
of cases, while the nuclei tend to have slightly lower FUV/H$\alpha$
ratios, suggesting either a harder ionization source or higher extinction. The slight deviation
from expected FUV/H$\alpha$ ratios for continuous star formation
(Leitherer \etal 1999) may be due to the fact that we have made no
attempt to correct for intrinsic extinction due to dust or due to a top-heavy
($\alpha \ll 2.35$) IMF. We note that modest amounts of dust
(E(B-V)~$\sim$~0.2) in the most dense regions of the ICM can account
for this deviation. Ideally, one would like spatially-resolved optical
spectra of the filaments in order to constrain the heat source and
intrinsic reddening of the filaments. We plan on addressing this issue
in upcoming studies. Comparing the estimates of the star formation
rates based on FUV, H$\alpha$ and MIR luminosities to the
spectroscopically-determined X-ray cooling rate suggests a star
formation efficiency of 14$^{+18}_{-8}$\%. This value represents the
time-averaged fraction, by mass, of gas cooling out of the ICM which
turns into stars and agrees well with the stars-to-gas fraction of $\sim$20--30\% required by simulations to reproduce the observed stellar mass function. This result provides a new constraint for studies of star formation in accreting systems. Many aspects of this simplified
scenario are still not well understood, including whether the star
formation is similar to that seen in nearby spirals or vastly
different.  We intend to investigate such differences via an assortment of star formation indicators from the UV to radio in future work.

\section*{Acknowledgements}
Support for this work was provided to M.M.\ and S.V.\ by NSF through
contracts AST 0606932 and 1009583, and by NASA through contract HST GO-1198001A.  We
thank E.\ Ostriker and A.\ Bolatto for useful discussions. We also thank the technical staff
at Las Campanas for their support during the ground-based
observations, particularly David Osip who helped in the commissioning
of MMTF.

\clearpage
\section*{Appendix}

\begin{figure*}[h]
\centering
\begin{tabular}{c}
\includegraphics[width=0.9\textwidth]{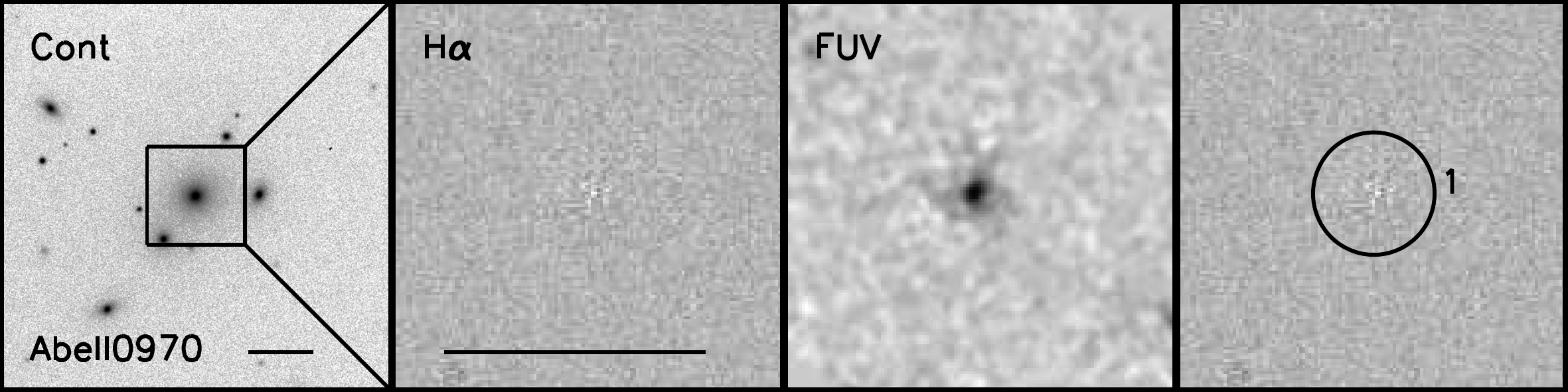}\\
\includegraphics[width=0.9\textwidth]{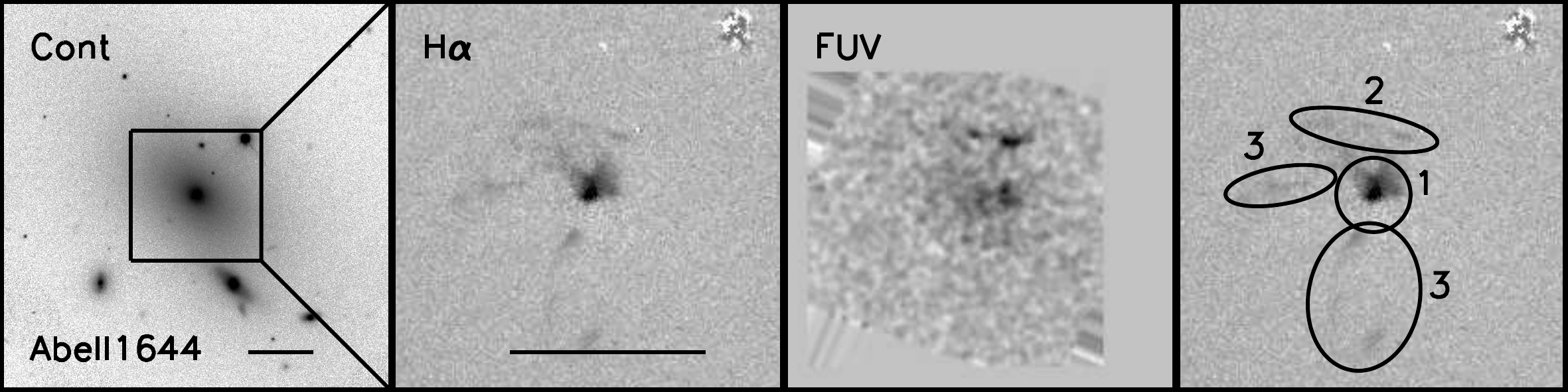}\\
\includegraphics[width=0.9\textwidth]{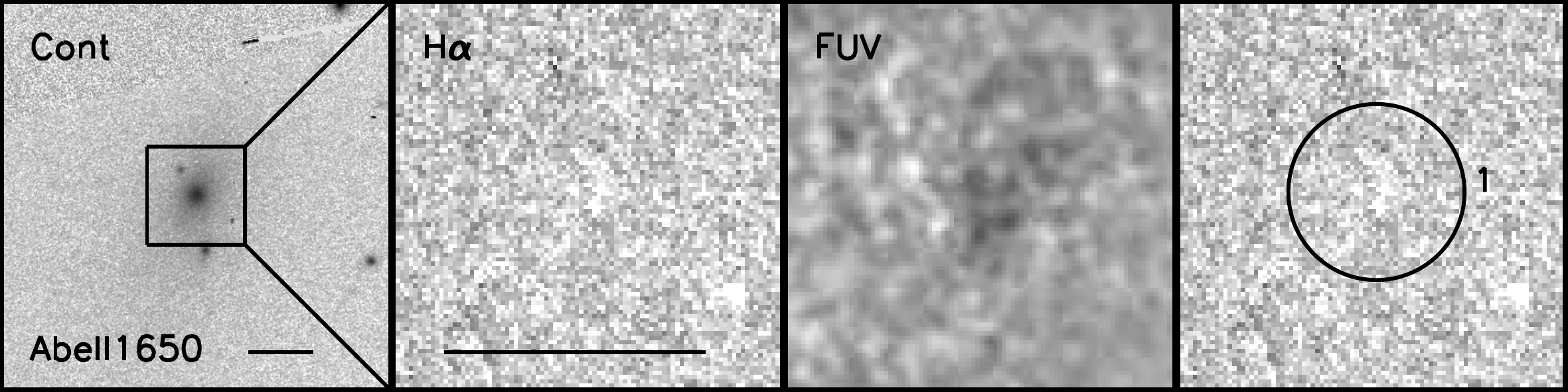}\\
\includegraphics[width=0.9\textwidth]{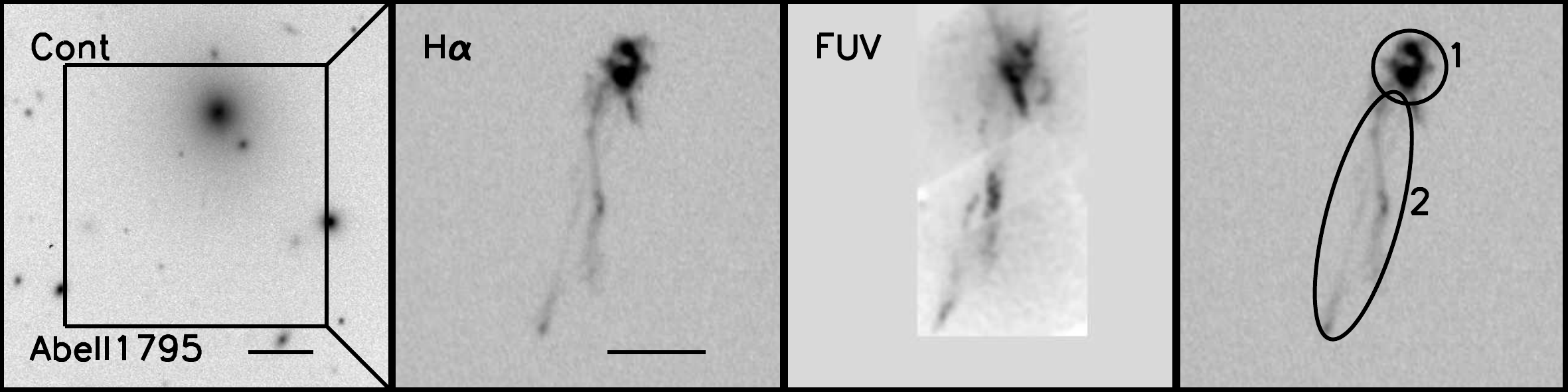}\\
\includegraphics[width=0.9\textwidth]{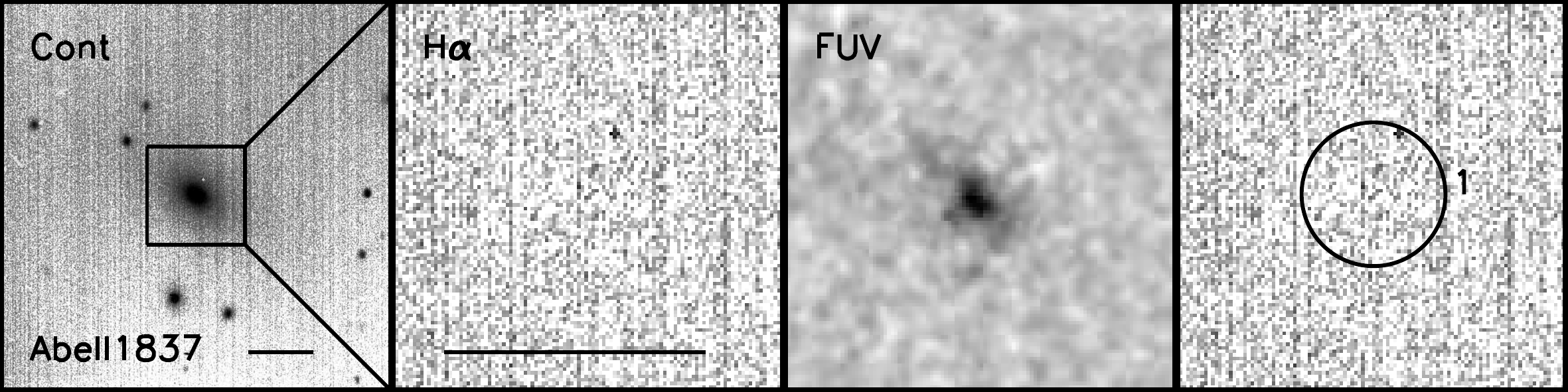}\\
\end{tabular}
\caption{Optical and FUV data for the 15 clusters in our
sample. From left to right the panels are: 1) MMTF red continuum
image, 2) MMTF continuum-subtracted H$\alpha$ image, 3) ACS/SBC FUV redleak-subtracted image (F150LP-F140LP), 4) H$\alpha$ image with extraction regions defined. The horizontal scale bar in the left two panels represents 20 kpc. The H$\alpha$ and FUV images are zoomed in relative to the red continuum image. The square region in the red continuum panels represents
the field of view for the zoomed-in panels. The grayscale in all
images is arbitrarily chosen in order to enhance any morphological
features.}
\label{bigfig}
\end{figure*}

\addtocounter{figure}{-1}
\begin{figure*}[p]
\centering
\begin{tabular}{c}
\includegraphics[width=0.9\textwidth]{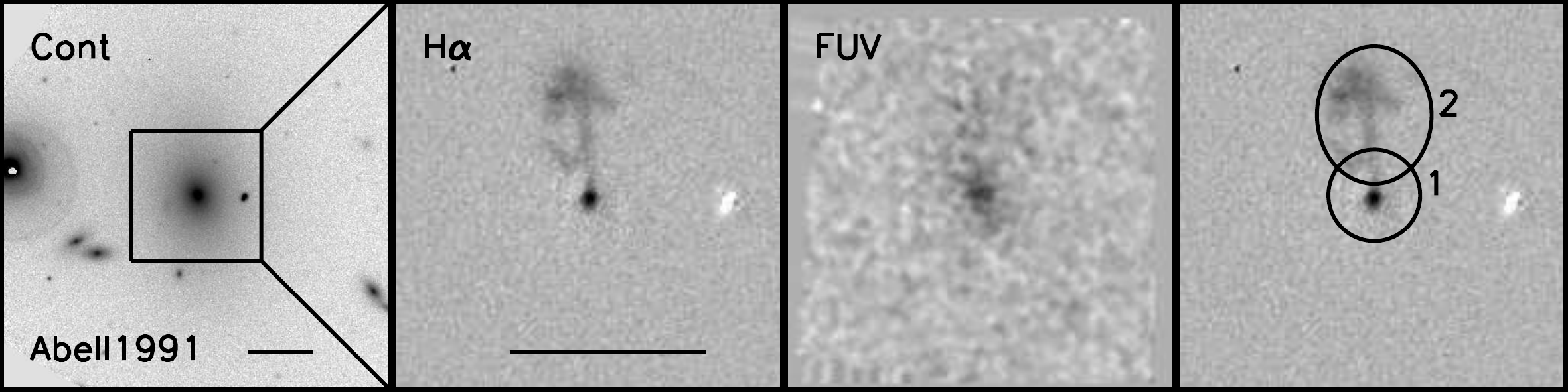}\\
\includegraphics[width=0.9\textwidth]{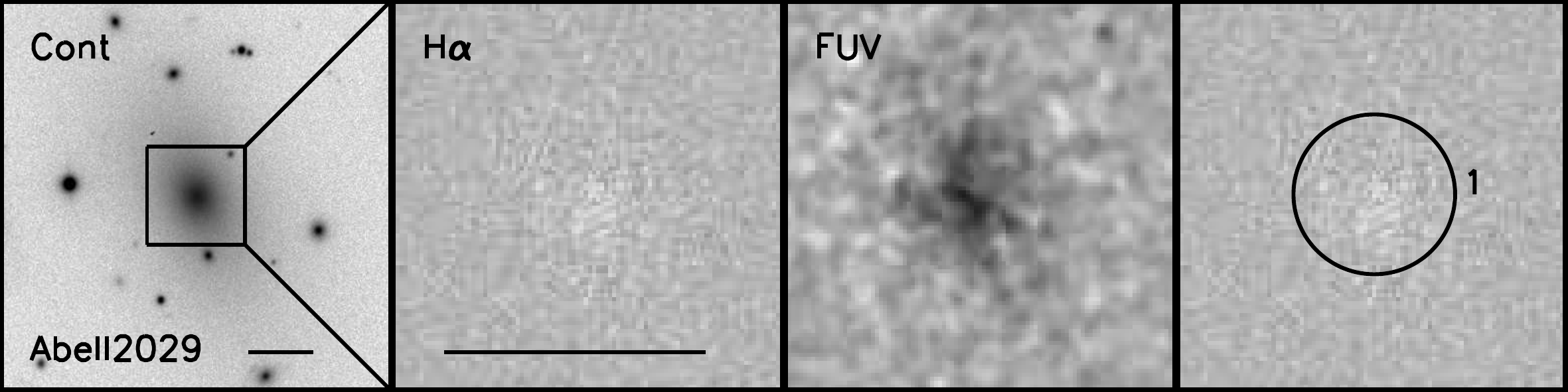}\\
\includegraphics[width=0.9\textwidth]{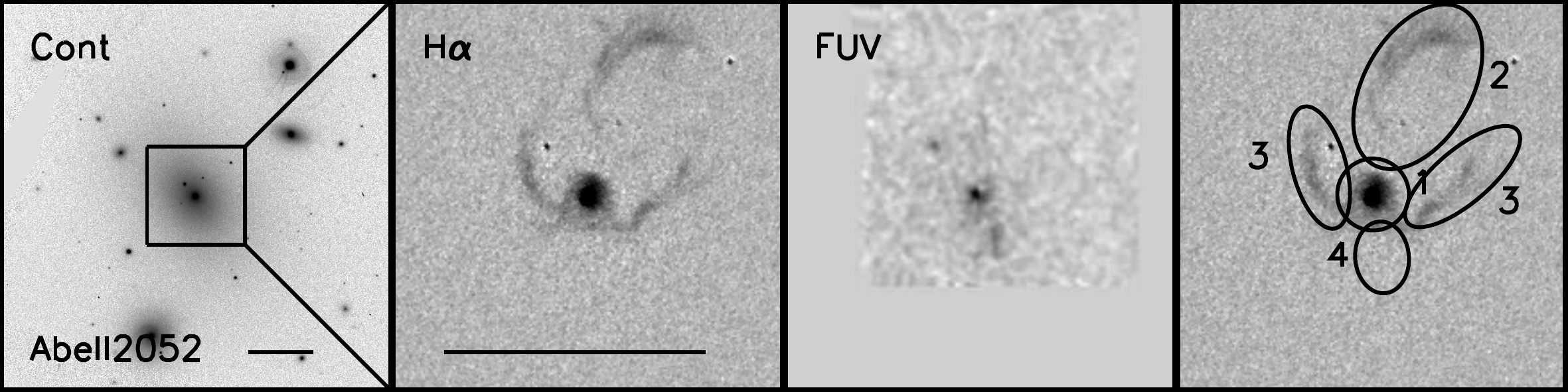}\\
\includegraphics[width=0.9\textwidth]{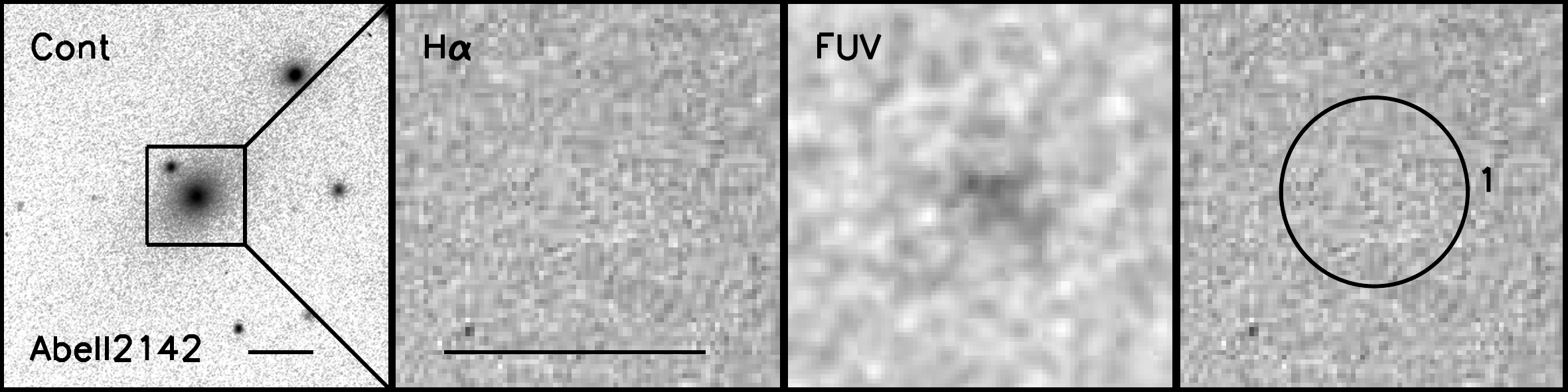}\\
\includegraphics[width=0.9\textwidth]{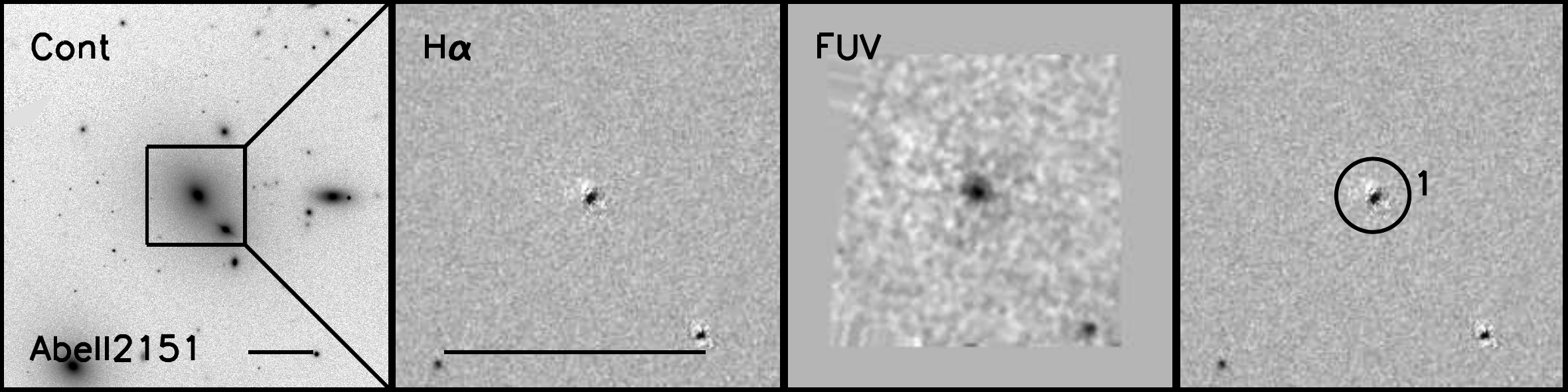}\\
\end{tabular}
\caption{Continued.}
\label{}
\end{figure*}

\addtocounter{figure}{-1}
\begin{figure*}[p]
\centering
\begin{tabular}{c}
\includegraphics[width=0.9\textwidth]{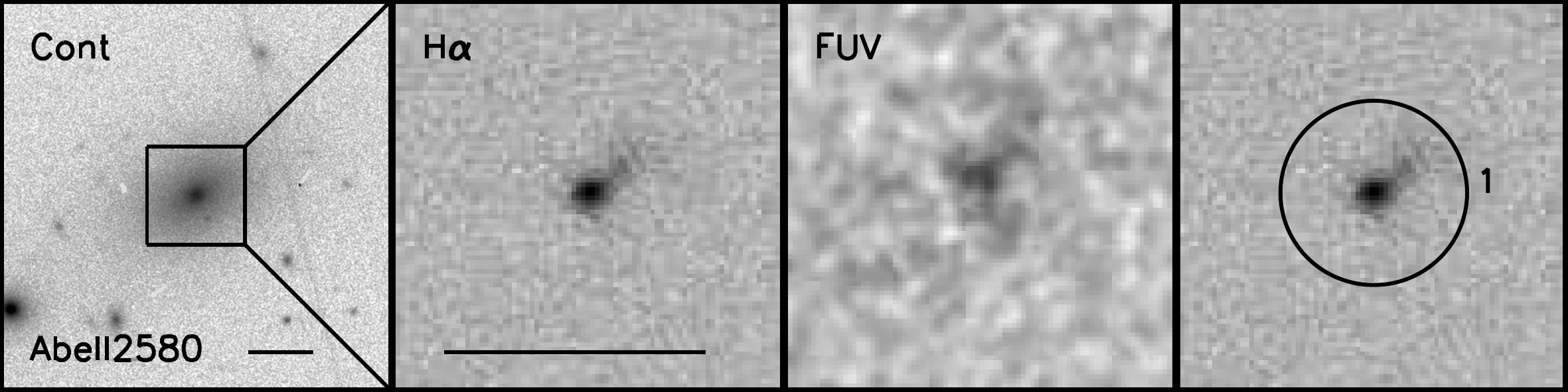}\\
\includegraphics[width=0.9\textwidth]{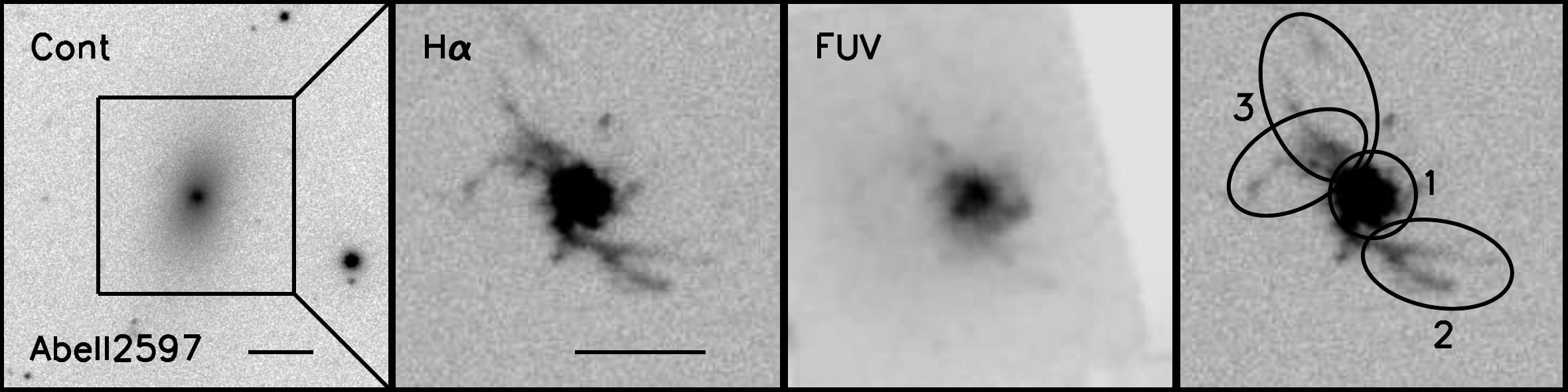}\\
\includegraphics[width=0.9\textwidth]{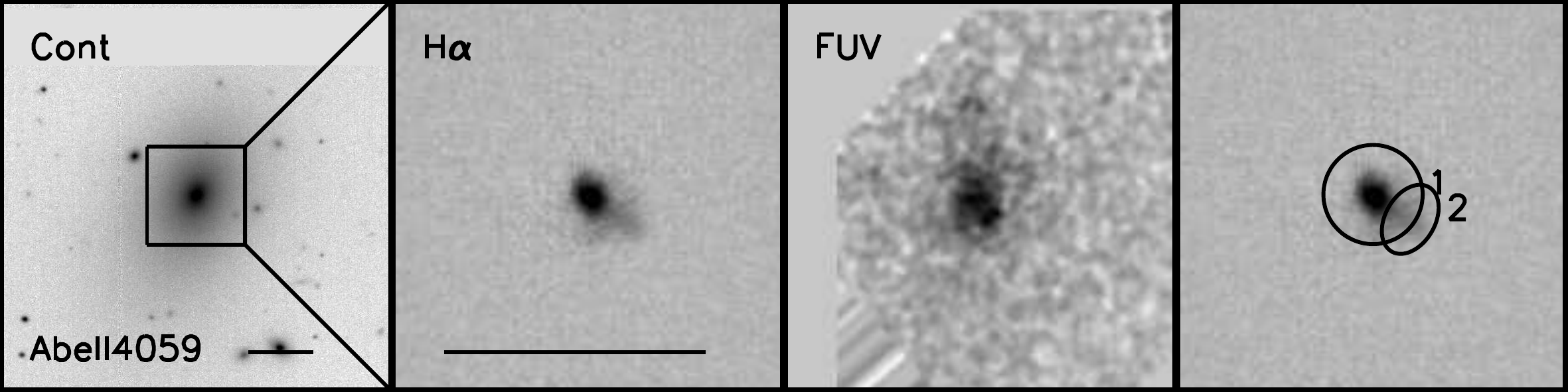}\\
\includegraphics[width=0.9\textwidth]{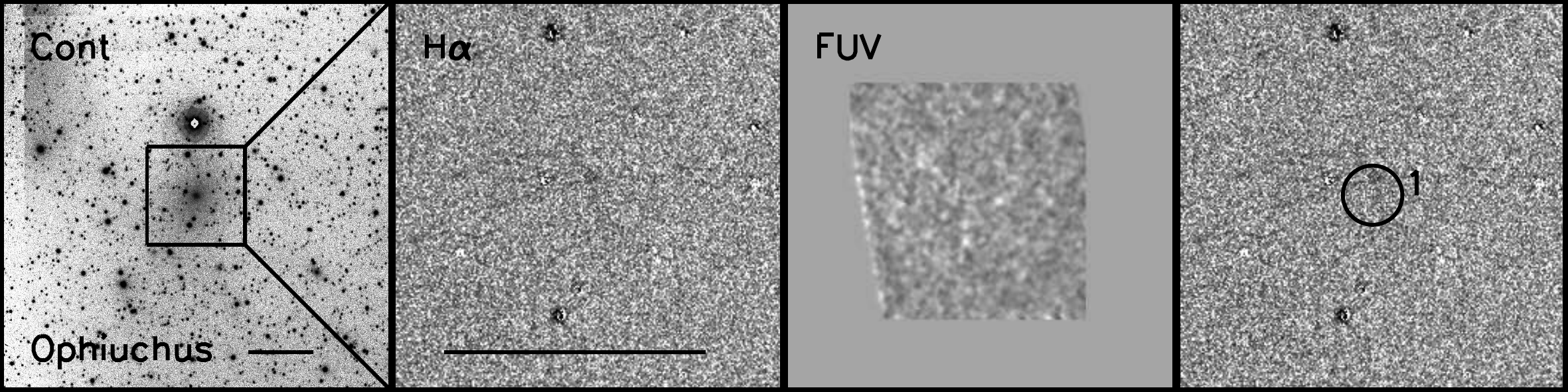}\\
\includegraphics[width=0.9\textwidth]{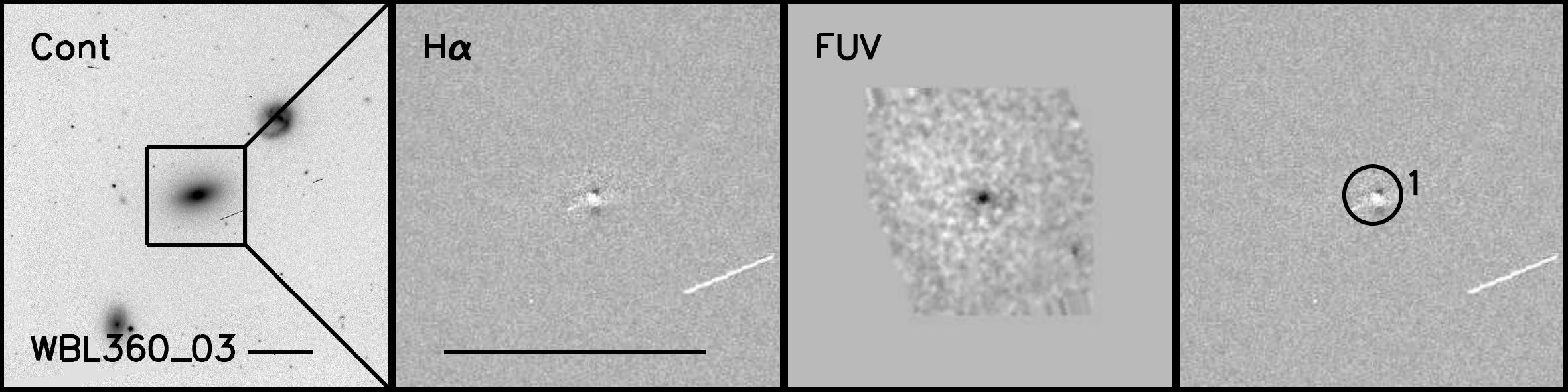}\\
\end{tabular}
\caption{Continued.}
\label{}
\end{figure*}

\end{document}